\documentclass[graybox]{svmult}

\usepackage{latexsym}
\usepackage{hyperref}
\usepackage{natbib}
\usepackage{lineno}
\usepackage{txfonts}
\usepackage{mathptmx}       
\usepackage{helvet}         
\usepackage{courier}        
\usepackage{type1cm}        
\usepackage[T1]{fontenc}
\usepackage{ae,aecompl}
\usepackage{makeidx}         
\usepackage[pdftex]{graphicx}        
\usepackage{multicol}        
\usepackage[bottom]{footmisc}

\makeindex             

\begin{document}

\title*{Asteroseismology of red giants \& galactic archaeology}
\author{Saskia Hekker}
\institute{Saskia Hekker \at Max Planck Institute for Solar System Research, Justus-von-Liebig Weg 3, 37077 G\"ottingen, DE\\
Stellar Astrophysics Centre, 120 Ny Munkegade, 8000 Aarhus, DK\\
 \email{hekker@mps.mpg.de}}
%
%
\maketitle

\abstract*{Red-giant stars are low- to intermediate-mass ($M \lesssim 10$~M$_{\odot}$) stars that have exhausted hydrogen in the core. These extended, cool and hence red stars are key targets for stellar evolution studies as well as galactic studies for several reasons: a) many stars go through a red-giant phase; b) red giants are intrinsically bright; c) large stellar internal structure changes as well as changes in surface chemical abundances take place over relatively short time; d) red-giant stars exhibit global intrinsic oscillations. \newline
Due to their large number and intrinsic brightness it is possible to observe many of these stars up to large distances. Furthermore, the global intrinsic oscillations provide a means to discern red-giant stars in the pre-helium core burning from the ones in the helium core burning phase and provide an estimate of stellar ages, a key ingredient for galactic studies. \newline
In this lecture I will first discuss some physical phenomena that play a role in red-giant stars and several phases of red-giant evolution. Then, I will provide some details about asteroseismology -- the study of the internal structure of stars through their intrinsic oscillations -- of red-giant stars. I will conclude by discussing galactic archaeology -- the study of the formation and evolution of the Milky Way by reconstructing its past from its current constituents -- and the role red-giant stars can play in that.}

\section{Introduction}
\label{sec:1}
Red-giant stars are low- to intermediate-mass ($M \lesssim 10$~M$_{\odot}$) stars that have exhausted hydrogen in the core. These extended, cool and hence red stars are key targets for stellar evolution studies as well as galactic studies for several reasons: a) many stars go through a red-giant phase; b) red giants are intrinsically bright; c) large stellar internal structure changes as well as changes in surface chemical abundances take place over relatively short time; d) red-giant stars exhibit global intrinsic oscillations. 

Due to their large number and intrinsic brightness it is possible to observe many of these stars up to large distances. Furthermore, the global intrinsic oscillations provide a means to discern red-giant stars in the pre-helium core burning from the ones in the helium core burning phase and provide an estimate of stellar ages, a key ingredient for galactic studies. 

In this lecture I will first discuss some physical phenomena that play a role in red-giant stars and several phases of red-giant evolution. Then, I will provide some details about asteroseismology -- the study of the internal structure of stars through their intrinsic oscillations -- of red-giant stars. I will conclude by discussing galactic archaeology -- the study of the formation and evolution of the Milky Way by reconstructing its past from its current constituents -- and the role red-giant stars can play in that.

The red-giant and asteroseismology part of this lecture is based on the lecture notes by Onno Pols\footnote{\url{https://www.astro.ru.nl/~onnop/education/stev\_utrecht\_notes/}}, the book by \citet{kippenhahn2012} and a review by \citet{hekker2016}. For more details I refer the reader to these sources. 

\section{Red-giant stars}
\label{sec:2}
In this section, I will provide a brief overview of some physical phenomena that are important in red-giant stars followed by a description of the different stages in stellar evolution of red-giant stars. 

\subsection{Physical phenomena}

\subsubsection{Convection}
\label{sect:conv} 
There is a physical limit to the energy flux that can be transported by radiation through a specific medium. If the temperature gradient becomes too steep convection takes over as the primary means of energy transport  \citep[Schwarzschild criterion;][]{schwarzschild1906}.
The Schwarzschild criterion states that convection is activated once the radiative temperature gradient ($\nabla_{\rm rad}$) exceeds the adiabatic temperature gradient ($\nabla_{\rm ad}$):
\begin{equation}
\nabla_{\rm rad} < \nabla_{\rm ad} \;,
\label{eq:schwarzschild}
\end{equation}
with 
\begin{equation}
\nabla_{\rm rad} = \left (\frac{\partial \log T}{\partial \log P} \right)_{\rm rad} = \frac{3}{16\pi a c G}\frac{\kappa l P}{mT^4} \,.
\end{equation}
describing the logarithmic variation of temperature $T$ with depth (expressed in pressure $P$) for a star in hydrostatic equilibrium in case energy is transported by radiation. Here, $a$ is the radiation constant, $c$ is the speed of light, $G$ is the gravitational constant, $\kappa$ is the opacity, $l$ is the local luminosity and $m$ is the mass coordinate, i.e. represents the mass contained inside a spherical shell of radius $r$.
The adiabatic temperature gradient is defined as:
\begin{equation}
\nabla_{\rm ad} = \left (\frac{\partial \log T}{\partial \log P} \right)_{\rm ad} \,.
\end{equation}

An alternative to the Schwarzschild criterion is the Ledoux criterion \citep{ledoux1947} which, in addition to the temperature gradients, takes into account the spatial variation of the mean molecular weight $\mu$. For an ideal gas this takes the form:
\begin{equation}
\nabla_{\rm rad} < \nabla_{\rm ad} + \nabla_{\mu} \;,
\label{eq:ledoux}
\end{equation}
with
\begin{equation}
\nabla_{\mu} = \left (\frac{\partial \log \mu}{\partial \log P} \right) \,.
\end{equation}
Note that $\nabla_{\rm rad}$ and $\nabla_{\mu}$ are spatial gradients, while $\nabla_{\rm ad}$ represents the temperature gradient in a gas element that undergoes a pressure variation.

Due to the nuclear reactions in the deep regions of a star  $\nabla_{\mu}$ is generally positive throughout the star. Hence the right-hand side of Eq.~\ref{eq:ledoux} takes larger values than the right-hand side of Eq.~\ref{eq:schwarzschild} and thus the mean molecular weight gradient incorporated in the Ledoux criterion has a stabilising effect.

\paragraph{\bf  Semi-convection}
In the regions that are convective according to the Schwarzschild criterion and stable according to the Ledoux criterion the true behaviour of the material remains unclear \citep{gabriel2014}. However in stellar evolution codes some form of `semi-convection' is often applied at the edge of a convective core. Semi-convection is a form of slow convection in which mixing takes place that is necessary to match observational constraints \citep[e.g.][]{lattanzio1983,langer1985}.

\paragraph{\bf Mixing length parametrisation}
Convection takes place over a large range of length scales which makes it complicated and expensive to model. To include convection in stellar modelling the mixing-length approximation is often used. The mixing length model was first proposed in 1925 by Ludwig Prandtl as a rough approximation of the distance or characteristic length a fluid parcel can travel before mixing with the surrounding fluid.
In stellar structure the most commonly used implementations of the mixing length $l_{\rm m}$ is from \citet{bohm-vitense1958}: 
\begin{equation}
l_{\rm m}=\alpha_{\rm MLT}\, H_P \,,
\end{equation}
with $H_P$ the pressure scale height, which is the radial distance over which the pressure changes by an e-folding factor:
\begin{equation}
H_P=\left| \frac{dr}{d \ln P}\right| = \frac{P}{\rho g}\,,
\label{eq:psh}
\end{equation}
where $\rho$ represents density and $g$ represents gravity. Note that the mixing length parameter $\alpha_{\rm MLT}$ is ordinarily calibrated to the Sun taking on a value between 1.2 and 2.2 depending on the stellar evolution code and choice of included physics. Although the solar value is often used as a fixed value in models of other stars there are indications that the value of $\alpha_{\rm MLT}$ should change as a function of evolution \citep[e.g.][]{trampedach2014}.

An alternative to the
mixing-length formalism,
described here,
is the approach by e.g. \citet{canuto1996} whom devised a full spectrum of turbulence which considers convection on different length scales.

\paragraph{\bf Convective overshoot}
The border between a radiative and convective layer may be soft in the sense that material on the convective side that approaches the boundary of stability with momentum penetrates into the radiative layer. This process that is referred to as convective overshoot, extends the convective region. In case of a convective core, convective overshoot can bring fresh fuel into the core prolonging the ongoing core burning phase. The extent of convective overshoot $l_{\rm ov}$ can be expressed as a fraction $\alpha_{\rm ov}$ of a local pressure scale height (Eq.~\ref{eq:psh}):
\begin{equation}
l_{\rm ov}=\alpha_{\rm ov}H_P \,.
\end{equation}
where $\alpha_{\rm ov}$ is a free parameter that can be calibrated against observations; typically $\alpha_{\rm ov} < 0.25$. 

\subsubsection{Electron degeneracy}
\label{sect:deg}
In very dense regions, such as the cores of low-mass (sub)giant stars ($M \lesssim 1.1$~M$_{\odot}$) or white dwarfs, the electron density is high enough to become degenerate. The density of electrons is described by Fermi-Dirac statistics as electrons are fermions with two spin states. Due to the Pauli exclusion principle that states that `two identical fermions cannot occupy the same quantum state' fermions will be forced to higher momentum states when the fermion density increases above the number of quantum states. The maximum number density of electrons $n_{e,\rm max}$ with momentum $p$ allowed by quantum mechanics is:
\begin{equation}
n_{e,\rm max}(p) = \frac{8 \pi p^2}{h^3} \,,
\end{equation}
with $h$ the Planck constant.
Hence, in very dense regions electrons have high momenta. The velocities of  these electrons exerts a higher pressure than inferred from classical physics. This is called degeneracy pressure which is nearly independent of temperature.

The transition between the classical ideal gas situation and a state of strong degeneracy occurs smoothly, and is known as partial degeneracy. Partial degeneracy has to be taken into account when 
\begin{equation}
n_e \gtrsim \frac{2(2 \pi m_e k T)^{3/2}}{h^3} \,,
\label{eq:ne}
\end{equation}
with $m_e$ the electron mass and $k$ the Boltzmann constant. The limit of strong (almost complete) degeneracy is reached when the electron density $n_e$ is roughly a factor 10 higher then the right hand side of Eq.~\ref{eq:ne}.

\subsubsection{Mirror principle}
At shell-burning regions, such as the hydrogen-shell burning region in red-giant stars, it is commonly seen that the region enclosed by the burning shell contracts, while at the same time the region outside the shell expands or vice versa when the region inside the burning shell expands, the region outside the shell contracts. This is referred to as the mirror principle. This is not a physical law as such, but an empirical observation, supported by the results of numerical simulations. 

The core contraction and envelope expansion of a star ascending the red-giant branch (i.e. evolve from D to G along the evolutionary track in Fig.~\ref{fig:HRD}) as well as the core expansion and envelope contraction while descending the red-giant branch (i.e. evolve from G to H along the evolutionary track in Fig.~\ref{fig:HRD}) are examples of the mirror principle.

\subsection{Different stages of evolution of red-giant stars}
\begin{figure}
\centering
\begin{minipage}{\linewidth}
\centering
\includegraphics[width=\linewidth]{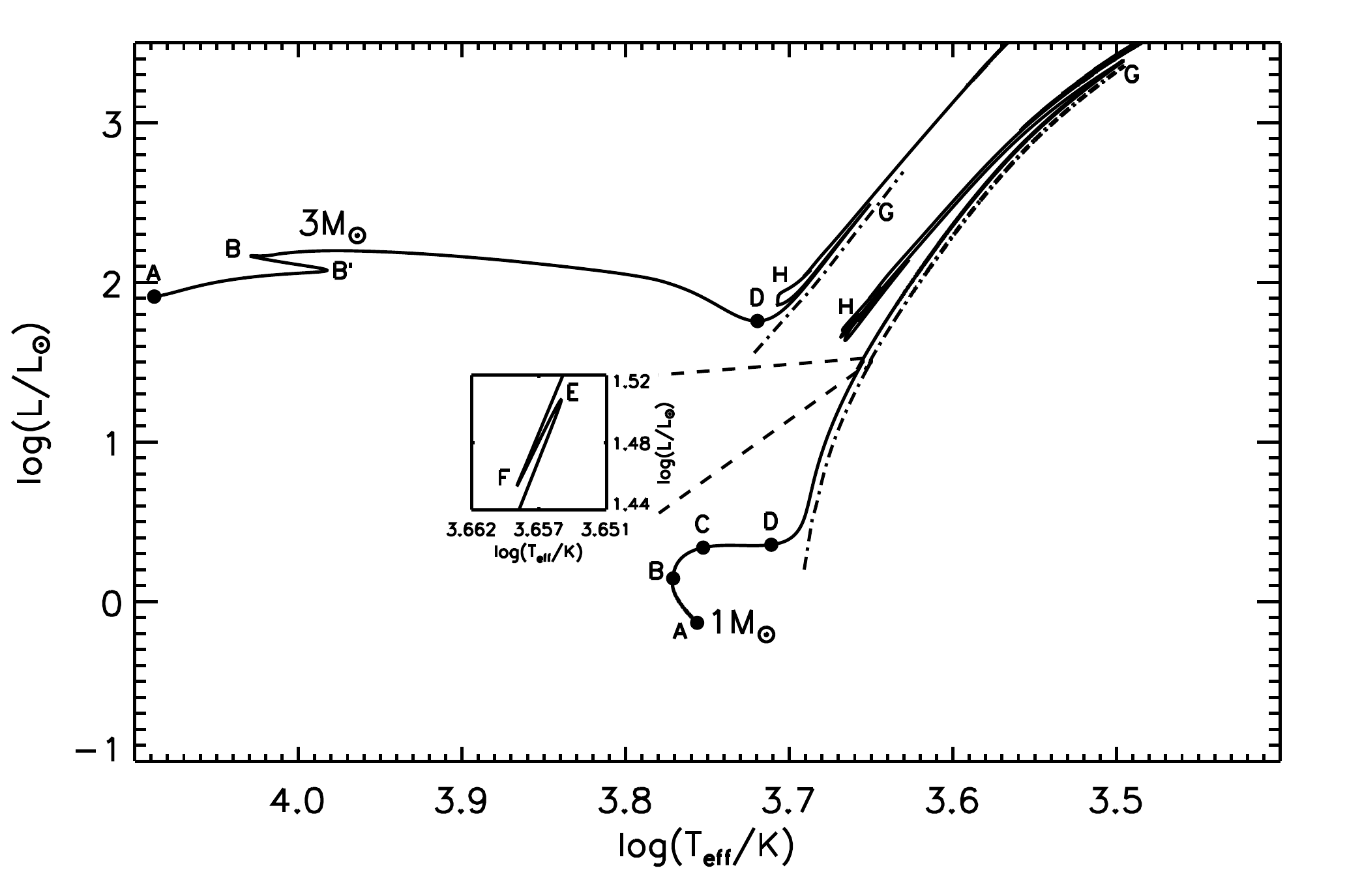}
\end{minipage}
\caption{Hertzsprung-Russell diagram of a 1~M$_{\odot}$ and a 3~M$_{\odot}$ evolutionary track. The inset shows the luminosity bump (see Section~\ref{sect:bump}) of the 1~M$_{\odot}$ track. The stellar model tracks are computed using the MESA stellar evolution code \citep{paxton2011} with solar metallicity. The letters indicate different phases of evolution: A = zero-age main-sequence; B$'$ = core hydrogen mass fraction $\approx 0.05$, B = start of thick shell burning; C = maximum extent of thick shell burning (in mass); D = start of thin shell burning; E = maximum bump luminosity; F = minimum bump luminosity; G = tip of the red-giant branch; H = helium core burning. The dashed-dotted lines provide a schematic indication of the location of the Hayashi lines for the two models. The phases B-D, D-G and G-H are referred to as subgiant, ascending red-giant and descending red-giant branches, respectively.}
\label{fig:HRD}
\end{figure}

Evolutionary tracks of both a 1~M$_{\odot}$ and a 3~M$_{\odot}$ stellar model with solar composition are shown in Fig.~\ref{fig:HRD}. Red-giant stars burn hydrogen in a shell around an inert helium core during the subgiant phase (B-D in Fig.~\ref{fig:HRD}) and while ascending the red-giant branch (RGB, D-G in Fig.~\ref{fig:HRD}) along which their radius increases and their surface temperature decreases. After the onset of helium core burning the stars reduce their size and increase their surface temperature (G-H in Fig.~\ref{fig:HRD}) while they reside in the so-called red-clump or secondary-clump phase where helium burning takes place in the core surrounded by a hydrogen burning shell (CHeB stars, H in Fig~\ref{fig:HRD}).
Here, I discuss  different stages of the evolution of red-giant stars indicated with different letters in Fig.~\ref{fig:HRD} occurring in both low- and intermediate-mass stars.  

\subsubsection{Low vs. Intermediate mass stars}
\begin{figure}
\centering
\begin{minipage}{\linewidth}
\centering
\includegraphics[width=\linewidth]{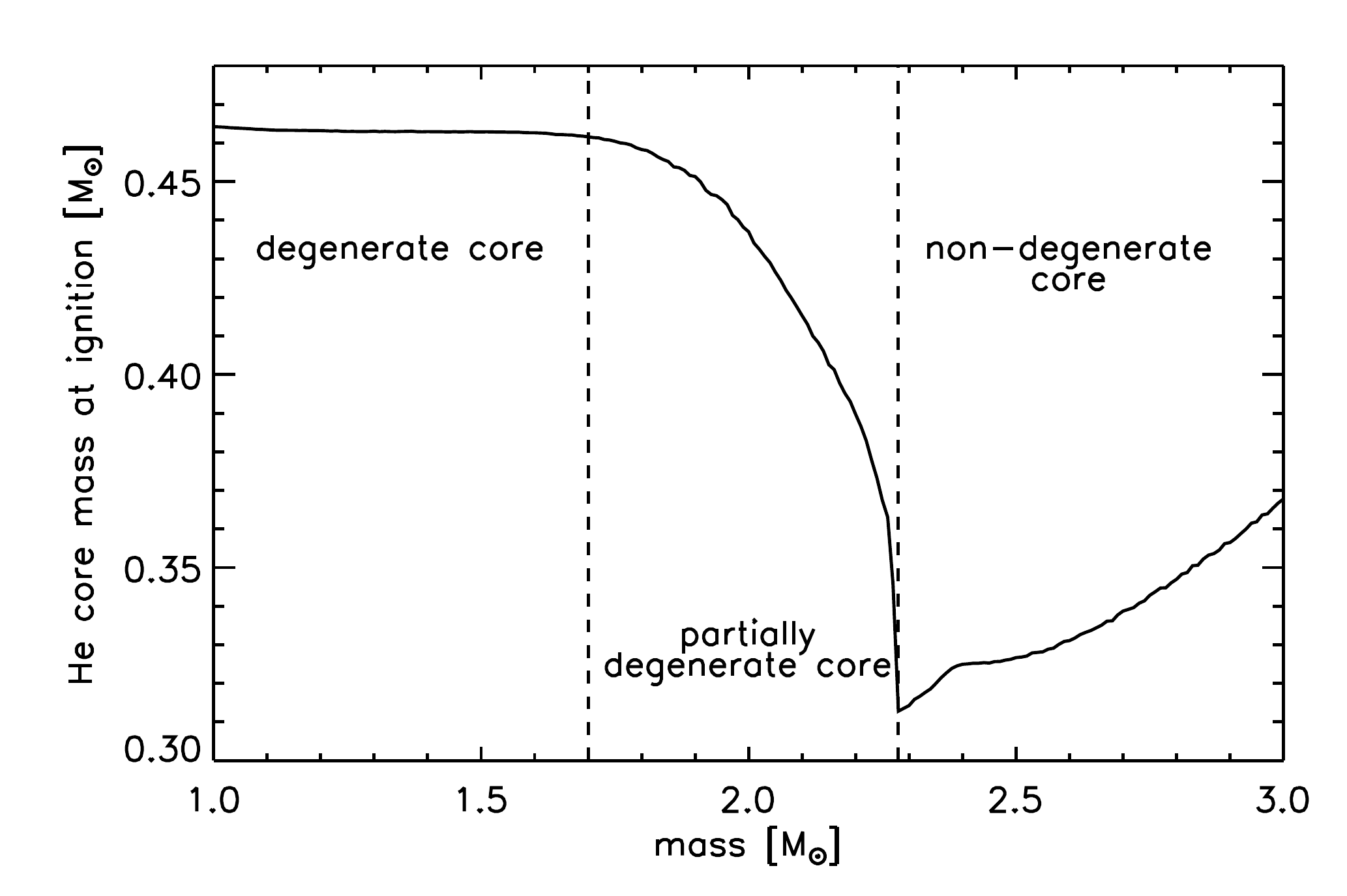}
\end{minipage}
\caption{Helium core mass at ignition vs. stellar mass for stellar models of solar metallicity computed with the MESA stellar evolution code \citep{paxton2011}. The vertical dashed lines are to guide the eye to the transitions in stellar mass of degenerate cores (right) to partially degenerate cores (centre) and non-degenerate cores (left). See Section~\ref{sect:deg} for a description of electron degeneracy.}
\label{fig:Hecoremass}
\end{figure}

The distinction whether a star is a low-mass star or an intermediate-mass star is based on the onset of helium core burning. For low-mass stars the inert helium core is degenerate on the RGB and helium ignition takes place under degenerate conditions (see Section~\ref{sect:deg} for more details about electron degeneracy). This occurs in stars with total masses between $\sim$0.48~M$_{\odot}$ and $\sim$2~M$_{\odot}$. These limits are defined by the lower limit of the critical mass needed to ignite helium and on the chemical composition of the star. The mass of the degenerate helium core is the same at ignition for these stars irrespective of the total mass (see Fig.~\ref{fig:Hecoremass}). 

Intermediate-mass stars do not develop a degenerate core and have a more gentle onset of helium burning. Hence, for these stars the helium core mass at ignition is a function of the total mass of the star. Fig.~\ref{fig:Hecoremass} shows the helium core mass at ignition as a function of total stellar mass.

\subsubsection{Hook}
In main-sequence stars (A-B in Fig.~\ref{fig:HRD}) above $\sim$1.1~M$_{\odot}$ the conditions in the core are such that a convective core is established. When hydrogen is nearly depleted (X$_{\rm core}$ $\approx$ 0.05, B$'$ in Fig.~\ref{fig:HRD}) the star contracts to maintain the energy production. This contraction leads to an increase in effective temperature and luminosity  until hydrogen is completely depleted in the centre. This results in a so-called ``hook" in the Hertzsprung-Russell diagram at the end of the main sequence indicated with B$'$-B in the 3~M$_{\odot}$ track in Fig.~\ref{fig:HRD}.

Note that for stars with masses below $\sim$1.1~M$_{\odot}$ no convective core develops on the main sequence and no `hook' is visible in the Hertzsprung-Russell diagram.

\subsubsection{Bottom of red-giant branch}
During the subgiant phase (B-D in Fig.~\ref{fig:HRD}) the stellar envelope expands and cools while a star evolves from the main sequence towards the Hayashi line.
The Hayashi line is the locus in the Hertzsprung-Russell diagram of fully convective stars, where a star cannot decrease its temperature further while maintaining hydrostatic equilibrium. The exact location of the Hayashi line is mass and metallicity dependent. The schematic location of the Hayashi line for a 1~M$_{\odot}$ and a 3~M$_{\odot}$ model with solar metallicity are indicated with the dashed-dotted lines in Fig.~\ref{fig:HRD}.
When approaching the Hayashi line a further increase of the radius causes an increase in luminosity (bottom of red-giant branch, D in Fig.~\ref{fig:HRD}), such that the star stays on the hot side of the Hayashi line with an extended convective envelope (see Section~\ref{sect:conv} for a description of convection). 

\subsubsection{First dredge-up}
\label{sect:dredge-up}
While the outer layers of the star expand after the exhaustion of hydrogen in the core, the convective envelope penetrates deep into the star to regions where the chemical composition has been altered by nuclear processes earlier in its evolution. The convection transports the chemical elements of these deep layers to the surface. This so-called first dredge-up changes the surface chemical abundances, for instance the $^{12}$C/$^{13}$C ratios are lowered. The use of `first' refers to the fact that more dredge-up episodes take place at later stages of stellar evolution. 

The first dredge-up occurs at the end of the subgiant phase and in the early red-giant phase, i.e. it starts on the 1~M$_{\odot}$ evolutionary track in Fig.~\ref{fig:HRD} between C and D and ends before E. For an intermediate-mass star the evolution between B and D is very fast (Hertzsprung gap) and the first dredge-up takes place in the early phase of the RGB (D-G in Fig.~\ref{fig:HRD}).
The end of the first dredge-up is when the convective region reaches a maximum depth in mass. Due to  the advance of the hydrogen-burning shell the convection recedes, leaving behind a chemical (mean molecular weight) discontinuity. The mean molecular weight in the mixed region is due to the combination of the pristine stellar abundance and the products of partial hydrogen burning. Therefore the mean molecular weight in this regions is lower than in the synthesised interior.  

\subsubsection{Luminosity bump}
\label{sect:bump}
Along the red-giant branch (D-G in Fig.~\ref{fig:HRD}) there is a clear zig-zag in the evolution path (E-F in Fig.~\ref{fig:HRD}) this is the so-called RGB luminosity bump. This temporal shift in luminosity and temperature happens when the hydrogen shell burning reaches the chemical discontinuity left behind by the first dredge-up at the deepest extent of the convective envelope. Naively, the decrease in the mean molecular weight at the chemical discontinuity causes the luminosity $L$ to decrease following $L\propto \mu^7M_{\rm core}^7$ \citep{refsdal1970}, where $M_{\rm core}$ is the core mass. However, \citet{jcd2015} showed that the situation is more complex with the luminosity beginning to decrease prior to the shell burning through the discontinuity. This is most likely caused by the fact that the decrease in $\mu$ outside the discontinuity starts affecting the hydrostatic structure just below the discontinuity, and hence the temperature, within and above the hydrogen-burning shell before it reaches the discontinuity.

The RGB luminosity bump is visible in stars up to a mass of about 2.2~M$_{\odot}$ as for more massive stars the hydrogen burning shell does not reach the chemical composition discontinuity before the onset of helium core burning.

\subsubsection{Onset of helium core burning}
For low-mass stars with a degenerate core the onset of helium burning happens in a very short episode: the so-called He-flash (G in the 1~M$_{\odot}$ track in Fig.~\ref{fig:HRD}). In the highly degenerate core the pressure does not depend on the temperature and therefore there is no thermostatic control (see Section~\ref{sect:deg}) to expand and cool the core. Therefore, at a temperature of about 10$^8$~K the onset of helium burning in degenerate conditions results in a thermal runaway process creating an enormous overproduction of nuclear energy during a very short time of order a few hours. This energy does not reach the stellar surface due to its absorption in non-degenerate layers. The onset of helium burning takes place at the location of maximum temperature that is generally not in the centre but in a concentric shell around the centre of the degenerate core. The temperature maximum is off centre because of extremely efficient neutrino losses in the core of electron degenerate material. Due to the off-centre ignition, it is predicted that the first main He-flash is followed by a series of subflashes till the degeneracy of the core is completely lifted and is back in equilibrium with helium burning in a convective core. Due to the degenerate state of the core prior to the He-flash the core mass at ignition does not depend on the total stellar mass (see Fig.~\ref{fig:Hecoremass}).

For intermediate-mass stars with non-degenerate cores the pressure and temperature in the core are related. This thermostatic feedback allows for gentle ignition of helium in the core of these stars. In this case the luminosity at which helium ignites is a function of the stellar mass (see Fig.~\ref{fig:Hecoremass}).

\subsubsection{Helium core burning}
After the onset of helium burning the luminosity decreases and the surface temperature increases, i.e., the star descends the red-giant branch (G-H in Fig.~\ref{fig:HRD}). The drop in luminosity is due to the lower energy generation in the hydrogen shell burning layer due to its decreased density and temperature, while the increased surface temperature is caused by the mirror principle increasing the core radius and decreasing the total stellar radius. The descend of the red-giant branch is rapid, while returning to central fusion results in a quiescent long-lived phase (the helium core burning phase, H in Fig.~\ref{fig:HRD}). In this phase a star has two energy sources: helium burning in the core (producing carbon and oxygen) and hydrogen burning (producing helium) in a shell around the core. 

Low-mass stars of different total masses have very similar helium core masses upon ignition of helium in the core. Therefore, these stars occupy a small region in a Hertzsprung-Russell diagram and stay at this position for a relatively long time: the red clump, i.e. at this position in a Hertzsprung-Russell diagram stars `clump' together when a whole population is observed. Within the red clump the luminosity is very tightly constraint by the helium core mass, while there exists some temperature dependence on the total stellar mass and composition.

For intermediate-mass stars the helium core mass at ignition is a function of stellar mass and therefore these stars do not reside in the red clump. Instead they form a secondary clump at lower luminosities and effective temperatures \citep{girardi1999}.

\section{Asteroseismology of red-giant stars}
\label{sec:3}
Asteroseismology is the study of the internal structure of stars through their intrinsic oscillations. In the outer layers of red-giant stars oscillations can be stochastically excited by the turbulent convection \citep[e.g.][]{goldreich1977,goldreich1988}. Effectively, some of the convective energy is transferred into energy of global oscillations. These type of oscillations are referred to as solar-like oscillations as they also occur in the Sun. For an extensive overview of solar-like oscillators I refer the reader to the review by \citet{chaplin2013}. In this lecture I focus on asteroseismic inferences of red giants that are of importance for galactic archaeology. For more details on asteroseismic inferences and the underlying theory I refer the reader to \citet{aerts2010} and the review by \citet{hekker2016}.

\begin{figure}
\centering
\begin{minipage}{\linewidth}
\centering
\includegraphics[width=\linewidth]{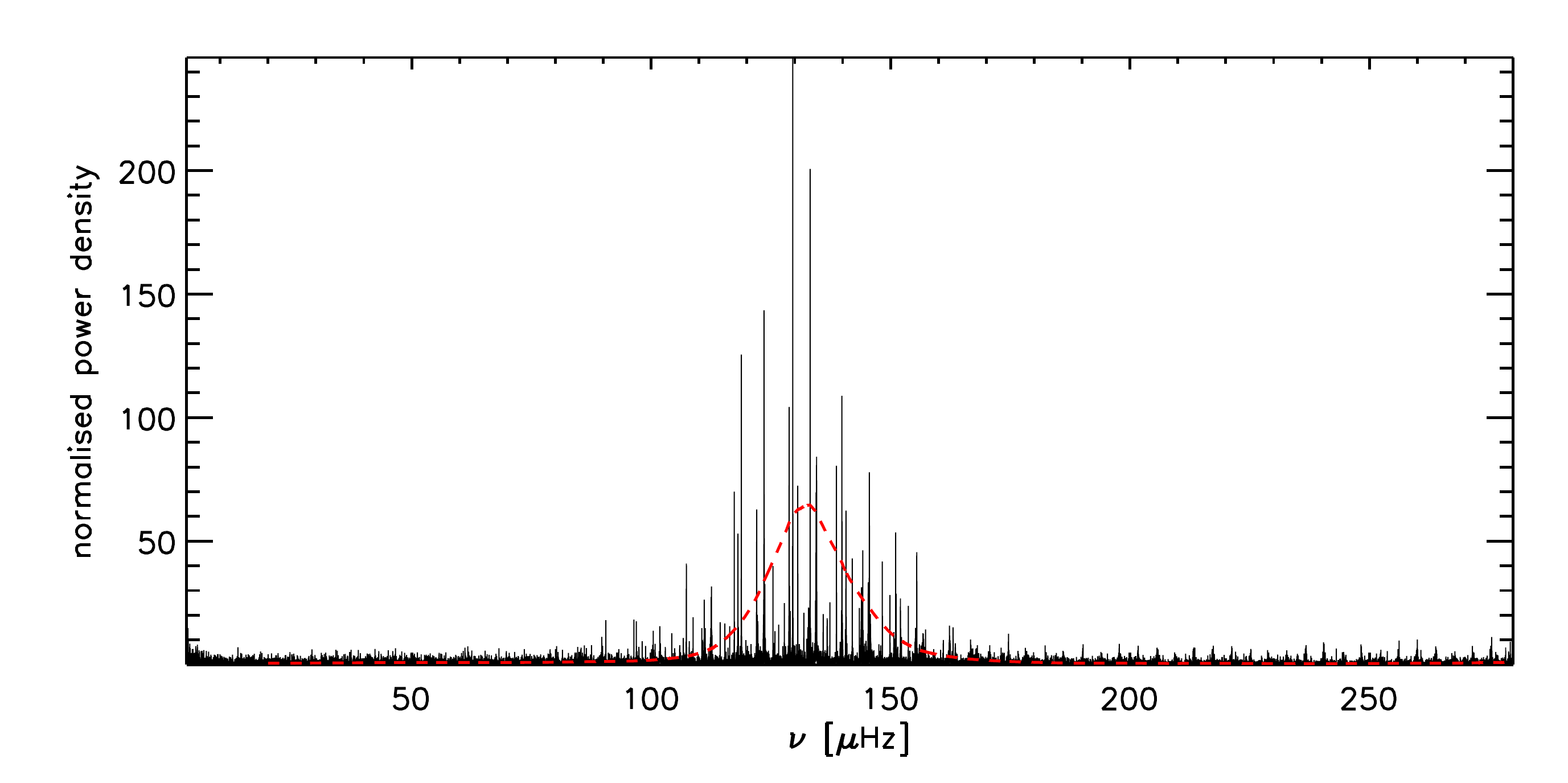}
\end{minipage}
\caption{Fourier spectrum normalised by the background signal of the red giant KIC~9145955. The red dashed curve is a heavily (triangular) smoothed power spectrum showing the power excess envelope of the oscillations. The amplitude of the smoothed power spectrum is enhanced for visual purposes. }
\label{fig:fourier}
\end{figure}

In stochastic oscillators essentially all modes are excited albeit with different amplitudes. This results in a clear oscillation power excess in the Fourier spectrum (see Fig.~\ref{fig:fourier}). Individual modes of oscillation in this power excess are described in terms of spherical harmonics. Hence the oscillations are described by their frequency $\nu$ and three quantum numbers: the radial order $n$ indicating the number of nodes in the radial direction, the spherical degree $l$ indicating the number of nodal lines on the surface, and the azimuthal order $m$ indicating the number of nodal lines that pass through the rotation axis.

\subsection{Oscillation pattern in Fourier space}
\label{sect:oscpat}
From the Fourier power spectrum in Fig.~\ref{fig:fourier} it is clear that oscillations reach observable amplitudes in a limited range in frequency with amplitudes that roughly follow a Gaussian shape. The centre of the oscillation power excess is referred to as $\nu_{\rm max}$, which is empirically found to be tightly related to the acoustic cut-off frequency $\nu_{\rm ac}$:
\begin{equation}
\nu_{\rm ac}=\frac{c}{4\pi H_p}
\label{eq:nu_ac}
\end{equation}
\citep[][using the approximation for an isothermal atmosphere]{lamb1932}, with $\nu_{\rm max} \approx 0.6 \nu_{\rm ac}$. The acoustic cut-off frequency is the maximum frequency of an acoustic eigenmode. At higher frequencies the waves are no longer trapped but traveling waves. 
It can be shown that $\nu_{\rm max}$ provides a direct measure of the surface gravity ($g$) when the effective temperature ($T_{\rm eff}$) is known \citep[e.g.][]{brown1991,kjeldsen1995}:
\begin{equation}
\nu_{\rm max} \propto \frac{g}{\sqrt{T_{\rm eff}}} \propto \frac{M}{R^2\sqrt{T_{\rm eff}}} \,,
\label{eq:numax}
\end{equation}
here $M$ and $R$ indicate the stellar mass and radius, respectively. \citet{belkacem2011} investigated the theoretical basis for this relation but a full explanation has not yet been found. 

\subsubsection{Acoustic modes}
\begin{figure}
\centering
\begin{minipage}{\linewidth}
\centering
\includegraphics[width=\linewidth]{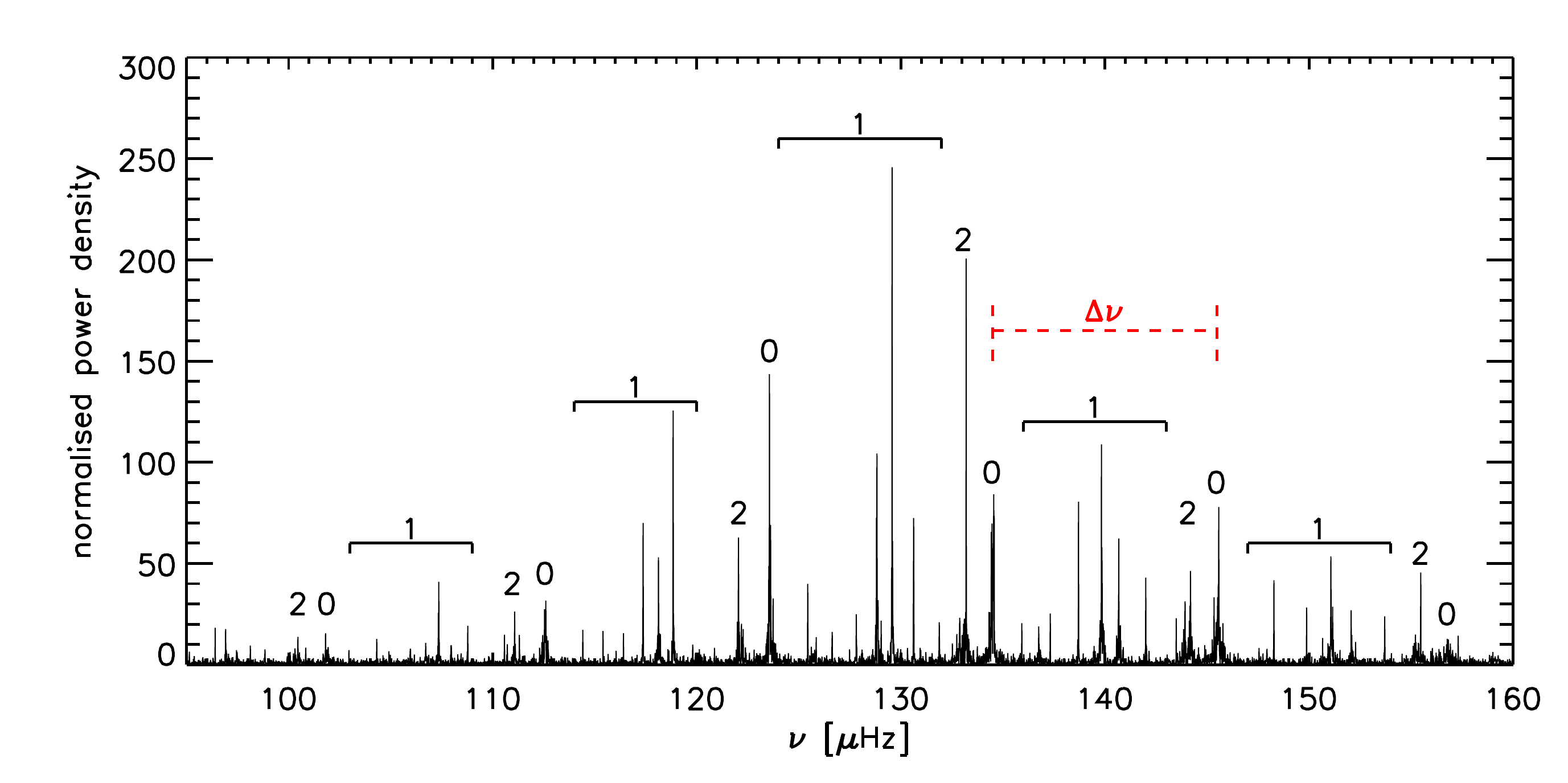}
\end{minipage}
\centering
\begin{minipage}{\linewidth}
\centering
\includegraphics[width=\linewidth]{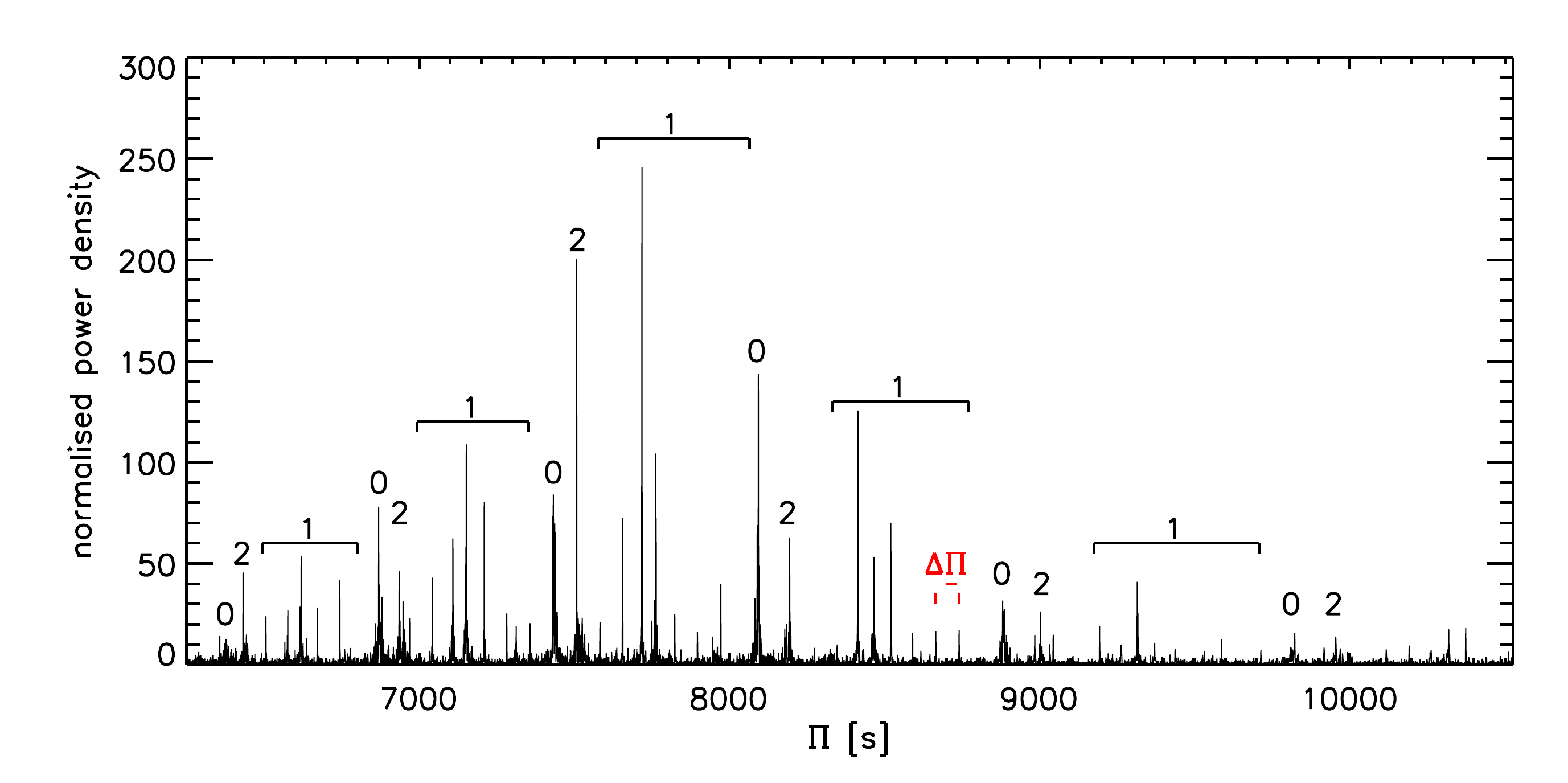}
\end{minipage}
\caption{Top: Fourier spectrum normalised by the background signal in the frequency range in which oscillations of the red giant KIC~9145955 are visible. The numbers indicate the degree $l$ of the modes. The large frequency between two consecutive radial ($l=0$) modes  is indicated in red. Bottom: same as top panel but now with period $\Pi$ in seconds on the x-axis and the observed period spacing $\Delta\Pi$ between two mixed dipole ($l=1$) modes indicted in red.}
\label{fig:fourier2}
\end{figure}

Following asymptotic theory \citep{tassoul1980}, high-order acoustic oscillation modes of solar-like oscillators with pressure as restoring force follow a near-regular pattern in frequency:
\begin{equation}
\nu_{n\,l} \simeq \Delta\nu(n+\frac{l}{2}+\epsilon)-d_{n\,l} \,,
\label{eq:tassoul}
\end{equation}
with $\Delta\nu$ the large frequency separation between modes of the same degree and consecutive radial order (see top panel of Fig.~\ref{fig:fourier2}), $\epsilon$ an offset and $d_{n\,l}$ a small correction to the leading order asymptotics, which is zero for $l = 0$.

$\Delta\nu$ is proportional to the inverse of the acoustic diameter, i.e. the sound travel time across a stellar diameter. Therefore, it can be shown that $\Delta\nu$ is a direct probe of the mean density ($\overline{\rho}$) of the star \citep{ulrich1986}:
\begin{equation} 
\Delta\nu = \nu_{n\,l} - \nu_{n-1\,l} = \left(2 \int_0^R \frac{{\rm d}r}{c} \right )^{-1} \propto \sqrt{ \overline{\rho}} \propto \sqrt{\frac{M}{R^3}},
\label{eq:dnu}
\end{equation}
with $r$ the distance to the centre of the star.

\subsubsection{Gravity modes}
\label{sect:mixed}
Following asymptotic theory \citep{tassoul1980}, high-order gravity modes with buoyancy as restoring force follow a near-regular pattern in period $\Pi_{n\,l}$:
\begin{equation}
\Pi_{n\,l} \simeq \Delta\Pi_l(n+\epsilon_g+1/2) \,,
\end{equation}
with $\epsilon_g$ a phase term and $\Delta\Pi_l$ period spacing (see bottom panel of Fig.~\ref{fig:fourier2}) defined as
\begin{equation}
\Delta\Pi_l = \frac{2\pi^2}{\sqrt{l(l+1)}} \left( \int_{r_1}^{r_2} N \frac{dr}{r}  \right)^{-1} \,,
\label{eq:DP}
\end{equation}
with $r_1$ and $r_2$ the turning points of the gravity mode cavity and $N$ the Bruntt-V\"as\"al\"a frequency:
\begin{equation}
N^2=g \left( \frac{1}{\Gamma_1} \frac{d \ln P}{dr} - \frac{d \ln \rho}{dr} \right) \,,
\label{eq:bruntt}
\end{equation}
where $g$ is the local gravitational acceleration, $P$ is pressure and $\Gamma_1=(\partial \ln P/ \partial \ln \rho)_{\rm ad}$. 
Note that $N^2$ is negative, and hence $N$ imaginary, in convectively unstable regions.

In red-giant stars pure gravity modes cannot be observed due to the extended convective envelope. However, in red giants the frequencies of the gravity modes in the core and the frequencies of the acoustic modes in the envelope have similar values such that resonant interactions between the modes allow for mixed acoustic-gravity nature to occur. Or put differently: in red giants essentially all non-radial ($l>0$) modes are mixed with different behaviour in different regions. The mixed mode frequencies are shifted by an amount depending on the coupling strength between the gravity and acoustic cavity. Due to these shifts the directly observed period spacing between consecutive mixed modes is smaller than the asymptotic value (Eq.~\ref{eq:DP}). The asymptotic $\Delta\Pi$ can be inferred from the observed period spacing \citep[e.g.][]{mosser2015}.

In practice mixed modes are mostly observed in dipole ($l=1$) modes  as for these modes the coupling between the pressure and gravity mode cavity is stronger and also because the period spacings are larger due to the dependence on $\sqrt{l(l+1)}$ (see Eq.~\ref{eq:DP}), and thus better resolved.

\subsection{Evolutionary state}
\label{sect:evolstate}
\begin{figure}
\centering
\begin{minipage}{\linewidth}
\centering
\includegraphics[width=\linewidth]{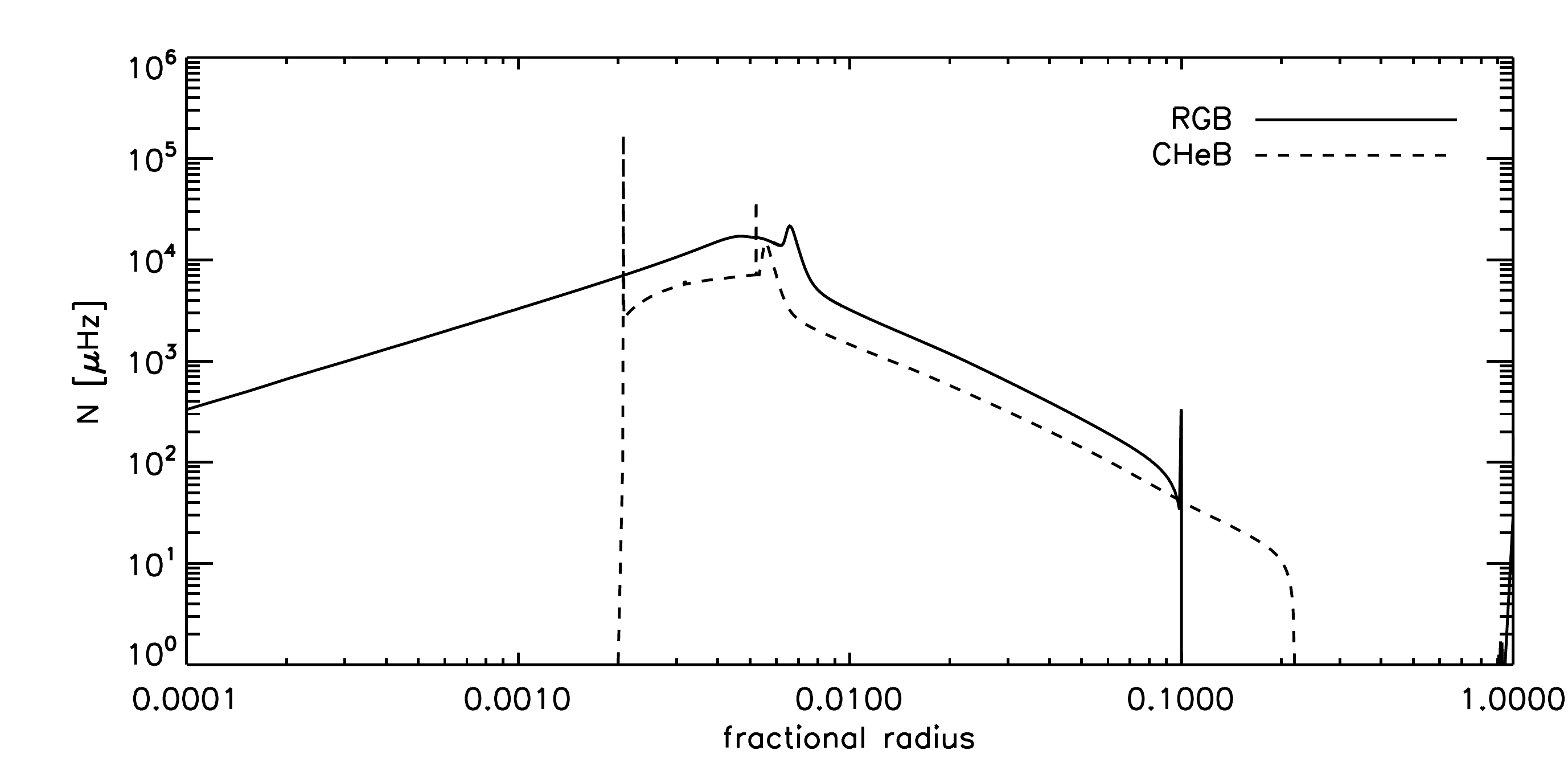}
\end{minipage}
\caption{Real part of the Bruntt-V\"as\"al\"a frequency $N$ of a 1~M$_{\odot}$ RGB model \citep[solid line, model 1 of][]{datta2015} and of a 1~M$_{\odot}$ CHeB model \citep[dashed line, black model in Fig. 10 of][]{constantino2015}.}
\label{fig:bruntt}
\end{figure}

Red-giant stars on the red-giant branch and in the helium core burning phase can occupy the same region in the Hertzsprung-Russsel diagram. Their surface properties can be very similar and hence it is difficult to distinguish between them based on classical observations such as surface temperature and brightness. However, mixed oscillation modes (Section~\ref{sect:mixed}) provide a means to probe the stellar cores where the difference between RGB and CHeB stars are significant. Stars on the RGB do not have convective cores while stars in the CHeB phase do have convective cores. Hence, the real part of the Bruntt-V\"as\"al\"a frequency (Eq.~\ref{eq:bruntt}) is zero in the convective core of CHeB stars and has a finite value in the core of RGB stars (see Fig.~\ref{fig:bruntt}). According to Eq.~\ref{eq:DP} this results in a significantly larger value of $\Delta\Pi$ for CHeB stars compared to RGB stars \citep[e.g.][and references therein]{bedding2011,jcd2014IAC,mosser2014}. Thus, asteroseismology provides a direct measure of the evolutionary phase of red-giant stars.

Alternatively it has been shown by \citet{kallinger2012} and \citet{jcd2014} that the differences in the core between
RGB and CHeB stars also cause differences in the thermodynamic state of the envelope. This results in a different location of the second helium-ionisation zone for RGB and CHeB stars. This causes a difference in the `local' phase term ($\epsilon$ in Eq.~\ref{eq:tassoul} when measured locally in a 2$\Delta\nu$ interval around $\nu_{\rm max}$) for stars in different evolutionary phases.

\subsection{Scaling relations}
Eqs~\ref{eq:numax} and \ref{eq:dnu} are often referred to as asteroseismic scaling relations:
\begin{equation}
\nu_{\rm max} = \nu_{\rm max,ref}\frac{M}{R^2\sqrt{T_{\rm eff}/T_{\rm eff, ref}}}
\label{eq:numaxs}
\end{equation}
\begin{equation} 
\Delta\nu = \Delta\nu_{\rm ref} \sqrt{\frac{M}{R^3}},
\label{eq:dnus}
\end{equation}
with $M$ and $R$ expressed in solar values and ``$_{\rm ref}$'' referring to reference values.
These scaling relations can be used to derive the stellar mass and radius. To do so a scaling with solar values as references has commonly been applied. This implicitly assumes that the Sun and the observed star have the same internal structure and only vary in size. From knowledge of stellar evolution it is known that this is not the case. Indeed, the high-quality observations of the CoRoT \citep{baglin2009} and \textit{Kepler} \citep{borucki2010} space telescopes as well as detailed analysis of models have shown that for Eq.~\ref{eq:dnus} this assumption is not entirely correct for stars with different properties, such as a different metallicity, or stars in different evolution phases. \citet{white2011,miglio2012,mosser2013,hekker2013,sharma2016,guggenberger2016}; Serenelli et al. (in prep.) have investigated the reference for the $\Delta\nu$ scaling relation (Eq.~\ref{eq:dnus}) in detail and generally propose either a correction to solar reference values \citep{white2011} or new reference values \citep[][Serenelli et al., in prep.]{mosser2013,hekker2013,sharma2016,guggenberger2016}. The correction by \citet{mosser2013} is based on the expanded asymptotic relation (Eq.~\ref{eq:tassoul}) known as the universal pattern \citep{mosser2011}. \citet{sharma2016} and Serenelli et al. (in prep.) performed an interpolation in a model grid to find a reference value for each model from individual frequencies, while \citet{guggenberger2016} used individual frequencies of models to derive a temperature and metallicity dependent reference function.

Although there is consensus in the community that for red-giant stars scaling to reference values that take the evolutionary phase, mass, metallicity and effective temperature into account is preferred over scaling to solar values, the exact reference value or function to be used is still a matter of debate. 

Note that for Eq.~\ref{eq:numax} a solar reference value is still commonly used as theoretical understanding of this relation is still insufficient to perform rigorous tests. Results for two red-giant binary stars show however that $\log(g)$ derived from the binary orbit and Eq.~\ref{eq:numax} are consistent (Theme\ss l et al. submitted), while stellar masses and radii derived using the orbits and asteroseismic scaling relations show discrepancies \citep[e.g.][Theme\ss l et al. submitted]{gaulme2016}.

\subsection{Comparison with stellar models}
\label{sect:GBM}
The scaling relations itself do not account for any knowledge we have about stellar evolution. To include this knowledge, it is also possible to compare observables of individual stars, e.g. \{$\Delta\nu$, $\nu_{\rm max}$, $T_{\rm eff}$, [Fe$/$H]\} or individual frequencies with models. This can be done in a grid-based modelling approach in which the observables are compared with a grid of stellar evolution models for which $\Delta\nu$ and $\nu_{\rm max}$ are computed using scaling relations (or alternatively in case of $\Delta\nu$ using individual frequencies). In this approach one does account for knowledge of stellar structure and evolution, as well as metallicity \citep[e.g.][]{gai2011}. An additional advantage is that it is possible in this way to obtain stellar ages in addition to stellar mass and radius, which is of importance for galactic archaeology.
Note that the results of grid-based modelling for red giants are generally more accurate when the evolutionary state, i.e. RGB or CHeB (see Section~\ref{sect:evolstate}), are known apriori.

In addition to grid-based modelling one can perform optimisation on a star by star basis \citep[e.g.][]{creevey2016}, although that is currently not feasible for large numbers of stars as required for galactic studies. Alternatively, one can use machine learning to obtain stellar parameters (including ages) in a fast and robust way \citep[][Angelou et al. submitted]{bellinger2016}.

\section{Galactic archaeology}
\label{sec:4}
Galactic archaeology is the study of the formation and evolution of the Milky Way galaxy by reconstructing its past from its current constituents. To this end it is important to know and understand the properties of the current constituents of the Milky Way, in terms of position, kinematics, chemical composition and age. These quantities contain information on how and where the stars were formed and hence provide the possibility to reconstruct the past.
As red-giant stars are common, intrinsically bright and show oscillations, these are prime targets to study the Milky Way. 


I will first provide a brief overview of the main components of the Milky Way galaxy and their formation scenarios followed by a discussion on some recent insights on the Milky Way obtained from red-giant stars. For more details on the Galaxy structure and its formation I refer the reader to reviews by \citet{freeman2002,rix2013,blandhawthorn2016} on which parts of this lecture are based. 

\subsection{Milky Way galaxy}
In this section I give brief descriptions of the main components of the Milky Way galaxy, i.e. the disk, the bulge, the halo and the dark matter halo. Star formation in the Milky Way has likely proceeded in phases, with limited overlap. Yet, the connection between the different components of the galaxy is far from understood \citep{allendeprieto2010}. However, N-body simulations such as the ones by e.g. \citet{athanassoula2016} are very promising and are currently able to simulate galaxies that have quantitatively the same components and shape as observed for the Milky Way.

\subsubsection{Disk}
\label{sect:disk}
The disk is a flat rotating radially extended part of the galaxy containing spiral arms. Commonly the disk is thought to consist of a thin and thick component. Different definitions can be used to separate these components: by looking only at the dynamical properties of stars or only at the chemical properties of stars. The properties of the resulting populations are however not identical. 
Nevertheless, in general the thick disk is more extended and contains old ($>$ 10 Gyrs), more metal poor stars ([Fe$/$H] between $-0.5$ and $-1$  dex) with larger enhancement of $\alpha$ elements compared to the thin disk (a chemically defined thick disk is therefore also commonly refer to as $\alpha$-rich disk). The $\alpha$ elements are elements for which the most abundant isotopes are integer multiples of four (the mass of the helium nucleus or $\alpha$-particle): stable $\alpha$-elements are carbon, oxygen, neon, magnesium, silicon, sulfur, argon and calcium.
The enhancement in [$\alpha/$Fe] is usually interpreted as evidence that the thick disk formed predominantly at times that were dominated by type II Super Novae (SNe), which have short-lived progenitors ($\sim$ few 10 Myrs) and are the main source of $\alpha$-elements. Additionally, thick disk stars show slow Galactic rotation and lag behind the thin disk rotation by roughly 50~km\,s$^{-1}$ with a dependence on the distance from the plane \citep{allendeprieto2010}. 

The thin disk stars are generally considered to be significantly younger than the thick disk stars and do not show enhancement in [$\alpha/$Fe]. The thin disk is less extended than the thick disk and shows a dependence on age with younger stars more confined to the mid-plane. 
Additionally, the velocity dispersions (random motions in three dimensions) increase with age. This is often referred to as `disk-heating'. This heating can occur due to gravitational scattering by objects (giant molecular clouds) or spiral density waves, or by collisions of satellite galaxies \citep{merrifield2001}.

Several authors \citep[e.g.][]{bovy2012,kawata2016} have argued that there is actually no thin disk/thick disk dichotomy and that the transition between thin and thick disk is rather a continuum of disks with an (abundance dependent) scale-length. 

A relation between the stellar age and the mean metallicity or between stellar age and velocity dispersion, i.e. age-metallicity relation (AMR) or age-velocity relation (AVR), would be fundamental observational input that could constrain the chemical and dynamical evolution of the galactic disk. However, there is much discussion about the existence of such relations \citep[e.g.][]{freeman2012,bergemann2014,kumamoto2017} and it may be that such relations can only be measured for young stars in the solar neighbourhood.

\subsubsection{Bulge}
\begin{figure}
\centering
\begin{minipage}{\linewidth}
\centering
\includegraphics[width=\linewidth]{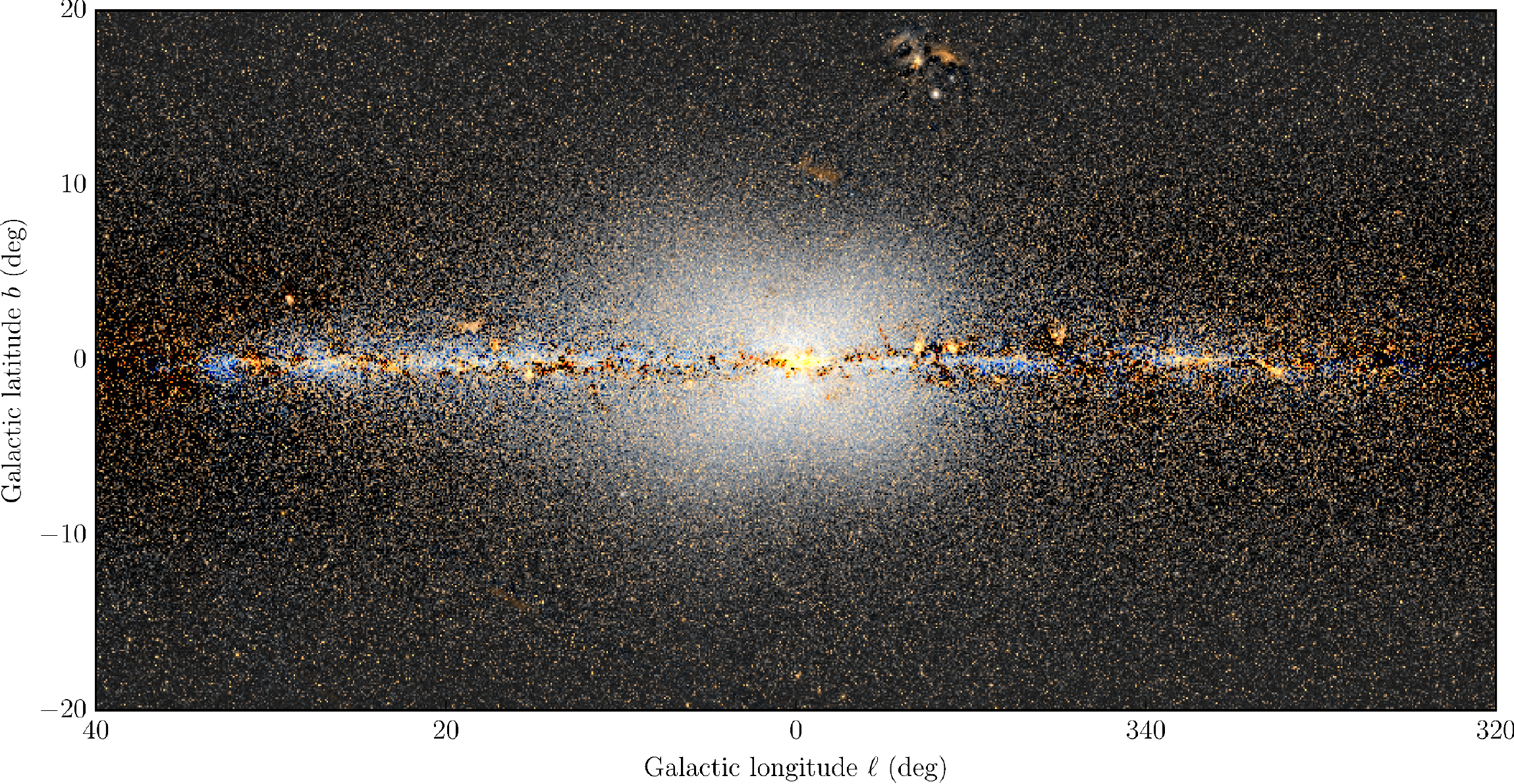}
\end{minipage}
\caption{WISE image of the Milky Way in W1 (3.4~$\mu$m) and W2 (4.6~$\mu$m) colour bands. An arcsinh stretch is used to allow the full dynamic range to be shown. Additionally, the median of each row of the image subtracted to provide a better contrast which reveals the X-shaped structure in better detail. Figure taken from \citet{nesslang2016}.}
\label{fig:bulge}
\end{figure}

The bulge is a densely populated part in the centre of the Milky Way with a boxy/peanut morphology and a X-shaped structure \citep[][and Fig.~\ref{fig:bulge}]{nesslang2016}, with its appearance depending on the angle it is viewed from and the stellar population that is looked at. A bulge can form naturally from the dynamics of a flat rotating disk of stars. If this is indeed happening the bulge formation takes 2-3 Gyrs to act after the disk has been formed. Hence in this scenario the bulge structure is younger than the bulge stars that were originally part of the inner disk \citep{freeman2012}.  

Another scenario could be that the bulge is formed out of mergers early in the formation of the galaxy. This is called a classical bulge and it is currently unclear whether the Milky Way contains such a bulge \citep{blandhawthorn2016}.

\subsubsection{Halo}
The halo is a sparsely populated extended (up to $\sim$100~kpc) part of the galaxy containing old metal poor stars. The halo stars have typically [Fe$/$H]~$<-1$~dex and ages greater than 12 Gyrs. The stellar density distribution of the halo follows a power law: $\rho \propto r^{-3.5}$ \citep{freeman2002}. The angular momentum of the halo is close to zero, while the velocity dispersion of the stars in three dimensions is large.

The galactic halo is formed at least partly through the accretion of small satellite galaxies that each carry their own signatures. Although accretion of dwarf galaxies is still taking place the most active accretion phase has probably ended about 12 Gyrs ago before the disk formation. Some (dynamical) traces of past accretion events are still present in the structure of the halo \citep[e.g.][]{helmi1999,grillmair2016}.

Globular clusters are collections of hundreds to thousands to millions of gravitationally bound stars that are an integral part of the stellar halo and orbit the galactic centre, both in the Milky Way as in other galaxies in the Local Group -- the group of galaxies to which the Milky Way belongs. Globular clusters may contain some of the first stars produced in the galaxy, although the origin and role of globular clusters in galactic evolution are still unclear.

\subsubsection{Dark matter halo}
The dark matter halo is only detected by its gravitational field and the existence is inferred through the effects on the motions of the stars and the gas in the galaxy. The mass of the dark matter halo is generally larger than the total mass of the visible components of a galaxy and extends well beyond the edge of the visible galaxy with a density that decreases farther from the galactic centre. Furthermore, dark matter does not seem to interact with other matter present in the galaxy. I note here that last year Erik Verlinde proposed a theory that shows that dark energy can explain the emerging gravity and that dark matter does not exist as a particle, but that dark matter is a phenomenon that emerges from dark energy \citep{verlinde2016}. 

\subsection{Chemodynamic model of the Milky Way}
State of the art models of the Milky Way are computed in the cosmological context and take the kinematics, chemistry and ages of stars into account in a self-consistent way \citep[e.g][]{minchev2013,minchev2014}. From these models it emerges naturally that stars can migrate in the  radial direction of the disk. This causes difficulties for galactic archeology in that stars can migrate significantly away from their place of birth \citep[e.g.][and references therein]{loebman2016}. This migration has to be taken into account in retracing the history of the stars in the Milky Way.

\subsection{Galactic archaeology and asteroseismology}
A key ingredient for galactic archaeology is stellar age. Stellar ages are difficult to derive as there is no observable that is sensitive to age and age only \citep{soderblom2010}. The classical way to derive stellar ages is by isochrone fitting using effective temperature ($T_{\rm eff}$), logarithmic surface gravity ($\log(g)$) and metallicity ([Fe$/$H]). This technique is very powerful for stars that belong to a cluster, but is limited in use for single field stars, due to degeneracies.

To determine stellar ages, spectroscopic information (mostly $T_{\rm eff}$ and ([Fe$/$H])) can be combined with asteroseismic information ($\Delta\nu$ and $\nu_{\rm max}$, see Section~\ref{sect:oscpat}) and stellar models in a grid-based modelling approach (see Section~\ref{sect:GBM}). The asteroseismic input (including the evolutionary phase in formation according for red giants, see Section~\ref{sect:evolstate}) provides further constraints on the models and therefore more precise age estimates. Further improvements could be made when individual oscillation frequencies are considered, but this is currently not feasible for a large number of stars as would be required for studies of the Milky Way.

The number of stars for which spectroscopic data and future parallaxes from Gaia \citep{perryman2001,lindegren2016} are and will be available is much larger than the number of stars with asteroseismic parameters as the latter require timeseries data. Ideally, the asteroseismic subsample can serve as a calibration set to determine ages of the larger set of stars. Indeed this seems possible as discussed in Section~\ref{sect:CN}.

\subsubsection{Asteroseismic calibration for spectroscopic samples}
\label{sect:CN}
In case there are features in stellar spectra that change as a function of age it is possible to transfer knowledge of stellar spectra of stars that have asteroseismic ages (or other information known accurately) to spectra that do not have asteroseismic information. Such a data-driven transfer is made by `The Cannon' \citep{ness2015} and allows to improve the scientific information obtained from spectra even if we do not understand the underlying reason of the correlation.

An example of age dependent spectroscopic features are the carbon (C) and nitrogen (N) abundances.
The first dredge-up (Section~\ref{sect:dredge-up}) causes a change in surface abundance as the stellar surface becomes mixed with material enriched in nitrogen and depleted in carbon, which causes a change in the ratio [C$/$N]. The value of the [C$/$N] abundance ratio depends on the CNO-processed material in the core at the end of the main sequence, and on the depth reached by the base of the convection zone, which both depend on the stellar mass: higher mass stars are comparatively richer in N and poorer in C with respect to lower mass stars. Additional mixing processes can subsequently act to change the [C$/$N] ratio at the upper red-giant branch.

\citet{martig2016} and \citet{ness2016} have used the chemical abundances and ratio as mentioned above together with asteroseismic masses and ages for a subset of stars to obtain an empirical link between mass and age on the one hand and C, N and [C$/$N] and [Fe$/$H] ratios on the other hand. Such an approach can potentially be used to calibrate relations that can be used on stars for which spectroscopic, but no asteroseismic data are available.

\subsubsection{$\alpha$-rich young stars}
Here, I provide an example where stellar ages derived using asteroseismology played an important role in studies of populations of stars in the Milky Way.

As described in Section~\ref{sect:disk} generally stars rich in $\alpha$-elements are formed at early times when the Milky way was dominated by type II SNe. Hence, $\alpha$-rich stars are expected to be old. \citet{chiappini2015,martig2015} did however detect a subsample of stars that are rich in $\alpha$-elements and at the same time are young based on asteroseismic age determinations. 

First studies by \citet{yong2016,jofre2016} show that these stars could be evolved blue stragglers, suggesting that the apparent young age is a consequence of a merger or mass transfer.

\section{Summary}
In this lecture I have focussed on red-giant stars and the role they can play in studying the Milky Way galaxy. These common, intrinsically bright oscillating stars allow us to probe the Milky Way till relatively large distances, and provide in specific fields improved precision on the obtained stellar parameters (e.g. ages) obtained by asteroseismology. The lack of accurate stellar age determinations for large numbers of stars is currently a limiting factor in galactic archaeology.

As it is technically feasible to take a spectrum for more stars than it is to take a timeseries as required for asteroseismology, techniques have been and are developed to use the more accurate and precise asteroseismic information and calibrate larger sets with these. One example of this is `The Cannon' \citep{ness2015}.

Additionally, the K2 mission \citep{howell2014} is currently still taking timeseries data for thousands of stars for the purpose of galactic archaeology \citep{stello2016}. In the near future the TESS \citep{ricker2014} and PLATO \citep{rauer2014} missions will also be launched. Although the prime aim of these missions is exo-planets and asteroseismology, the all-sky view of TESS and the step and stare mode of PLATO provide excellent prospects for galactic archaeology. These data combined with parallaxes from Gaia \citep{perryman2001,lindegren2016} and data from ground-based spectrographs will allow to improve our current picture of the Milky Way and from that provide clues about the formation and evolution of the Milky Way.

\begin{acknowledgement}
I would like to thank Alexey Mints, George Angelou and Maarten Mooij for useful discussions and comments on earlier versions of the manuscript. Furthermore, I acknowledge Earl Bellinger, Abishek Datta and Thomas Constantino for providing stellar evolution models used for figures. I acknowledge funding from the European Research Council under the European Community's Seventh Framework Programme (FP7/2007-2013) / ERC grant agreement no 338251 (StellarAges).

\end{acknowledgement}

\end{document}